\documentclass[twocolumn,english,prl,aps,floatfix]{revtex4}
\usepackage[T1]{fontenc}
\usepackage[latin9]{inputenc}
\usepackage[letterpaper]{geometry}
\usepackage{graphicx}
\usepackage{amssymb}

\makeatletter

\newcommand{\binom}[2]{{#1 \choose #2}}

\@ifundefined{textcolor}{}
{%
 \definecolor{BLACK}{gray}{0}
 \definecolor{WHITE}{gray}{1}
 \definecolor{RED}{rgb}{1,0,0}
 \definecolor{GREEN}{rgb}{0,1,0}
 \definecolor{BLUE}{rgb}{0,0,1}
 \definecolor{CYAN}{cmyk}{1,0,0,0}
 \definecolor{MAGENTA}{cmyk}{0,1,0,0}
 \definecolor{YELLOW}{cmyk}{0,0,1,0}
 }

\makeatother

\makeatother

\usepackage{babel}

\makeatother

\usepackage{babel}

\makeatother

\usepackage{babel}

\makeatother

\usepackage{babel}

\begin{document}

\title{Exact Zero Modes in Closed Systems of Interacting Fermions}

\author{G. Goldstein}

\affiliation{Physics Department, Harvard University, Cambridge MA 02138, USA}

\author{C. Chamon}

\affiliation{Physics Department, Boston University, Boston MA 02215 USA}
\begin{abstract}
We show that for closed finite sized systems with an odd number of
real fermionic modes, even in the presence of many-body interactions,
there are always at least two fermionic operators that commute with
the Hamiltonian. There is a zero mode corresponding to the total Majorana
operator, as shown by Akhmerov~\cite{Akhmerov}, as well as additional
linearly independent zero modes, one of which 1) is continuously connected
to the Majorana mode solution in the non-interacting limit, and 2)
is less prone to decoherence when the system is opened to contact
with an infinite bath. We also show that in the idealized situation
where there are two or more well separated zero modes each associated
with a finite number of interacting fermions at a localized vortex,
these modes have non-Abelian Ising statistics under braiding. Furthermore
the algebra of the zero mode operators makes them useful for fermionic
quantum computation \cite{key-2}. 
\end{abstract}
\maketitle

\section{Introduction}

Zero modes in non-interacting systems, \textit{i.e.} eigenstates annihilated
by a single-particle Hamiltonian, have a long history in physics and
in mathematics. Zero energy states are associated to certain types
of topological defects in the background fields in which electrons
or quasiparticles propagate. The first example of such modes in physics
appeared in the seminal work of Jackiw and Rebbi~\cite{Jackiw-Rebbi}
in one-dimensional and three-dimensional systems, where the topological
defects were domain walls and hedgehogs, respectively. In both these
examples the physical consequence of the zero modes is the fractionalization
of electron charge. Fractional charges can also be bound to vortices
in a Kékule dimerization pattern in two-dimensional graphene-like
systems~\cite{Hou-Chamon-Mudry}. The zero mode solutions in two-dimensions
were first found by Jackiw and Rossi~\cite{Jackiw-Rossi} in the
study of Dirac fermions in the background of scalar and vector gauge
fields of the Abelian Higgs model. In the condensed matter context
this corresponds to a superconductor (where charge cannot be fractionalized,
since it is not conserved). The number of zero modes in such system
of Dirac fermions in two-dimensions equals the magnitude of the net
vorticity independent of the details of the profile of the Higgs fields,
a result that was shown by Weinberg~\cite{Weinberg} to be tied to
the index theorem.

A modern example of a physical realization of the model in Ref.~\onlinecite{Jackiw-Rossi}
was presented by Fu and Kane~\cite{Fu-Kane}, who showed that a Dirac-type
matrix equation governs surface excitations in a topological insulator
in contact with an s-wave superconductor. A vortex in the superconducting
order parameter leads to a zero mode solution. Because of the reality
conditions imposed by the symmetries of the Bogoliubov-de Gennes (BdG)
equations describing the superconductor within the mean-field approximation,
the zero energy solutions correspond to Majorana zero modes, which
are the focus of our study. Majorana fermions are self-adjoint operators
$\gamma_{i}$ which can be written as a sum of an annihilation and
creation operator for one fermion mode and which satisfy the algebra:
\begin{equation}
\left\{ \gamma_{i},\,\gamma_{j}\right\} =2\delta_{ij},\,\gamma_{i}^{\dagger}=\gamma_{i}.\label{eq:BasicDefintion}\end{equation}
 Because they are zero modes of some mean field Hamiltonian, $\left[{H}_{\mathrm{MF}},\,\gamma_{i}\right]=0$,
these modes are in principle protected from decoherence as the mean
field Hamiltonian, when restricted to the subspace generated by these
modes, is zero. Recently it has been argued that quantum and classical
fluctuations in open infinite systems (for example when the system
is in contact to a bath) lead to decoherence of information stored
in such modes \cite{key-1}. Below, instead, we shall focus on closed,
finite systems, which have markedly different properties from those
coupled to an infinite environment.

The purpose of this letter is to study zero modes of interacting many-body
fermionic Hamiltonians, beyond mean-field approximations. We will
assume that the relevant degrees of freedom may be described by an
odd number of Majorana fermions $\left\{ \gamma_{1},\gamma_{2},\dots,\gamma_{2N+1}\right\} $.
This formalism also handles the case when complex fermions are present,
as we may change basis from complex to Majorana fermions: $c_{j}=\frac{1}{2}\left(\gamma_{2j}+i\gamma_{2j+1}\right),\: c_{j}^{\dagger}=\frac{1}{2}\left(\gamma_{2j}-i\gamma_{2j+1}\right)$.
For an interacting many-body Hamiltonian, a zero mode means a Hermitian
fermionic operator \begin{equation}
\mathcal{O}=\sum_{i}\alpha_{i}\;\gamma_{i}+i\sum_{i,j,k}\beta_{i,j,k}\;\gamma_{i}\gamma_{j}\gamma_{k}+\dots\;,\label{eq:zero-mode-form}\end{equation}
 written as a multinomial with sums and products of $\gamma_{i}$'s,
that commutes with the Hamiltonian, $\left[H,\mathcal{O}\right]=0$.
For any such operator, $\mathcal{O}$, $\exp\left(itH\right)\mathcal{O}\exp\left(-itH\right)=\mathcal{O}$
for all times $t$. As such there is no decoherence of the information
stored in the correlators of such operators.

We will find below, for systems of interacting fermions, $2^{N}$
linearly independent solutions of the form given in Eq.~(\ref{eq:zero-mode-form}).
We will also extend our results to the case when interactions include
bosonic modes (with finite dimensional Hilbert space) coupled to the
Majorana modes.

\section{Quadratic Hamiltonians}

Let us start, as a warm up, with the simplest case where $H^{\mathrm{Gauss}}=i\sum_{i,j}h_{i,j}\;\gamma_{i}\gamma_{j}$
with $h_{i,j}=-h_{j,i}$ and $h_{i,j}$ real. We note that any quadratic
Hamiltonian may be written in this manner. Generic eigenoperator solutions
satisfying $\left[H^{\mathrm{Gauss}},\mathcal{O}_{\lambda}\right]=\lambda\,\mathcal{O}_{\lambda}$
are obtained by computing the commutators for operators of the form
$\mathcal{O}=\sum_{i}\alpha_{i}\gamma_{i}$ using the relations Eq.~(\ref{eq:BasicDefintion}),
and matching the coefficients multiplying each operator $\gamma_{i}$
on both sides of the equation. One arrives in this manner at an eigenvalue
equation for the matrix \begin{equation}
{\mathcal{H}}^{\mathrm{Gauss}}=4i\left(\begin{array}{ccccc}
0 & h_{1,2} & h_{1,3} & \cdots & h_{1,2N+1}\\
h_{2,1} & 0 & \ddots &  & \vdots\\
h_{3,1} & \ddots & 0 &  & \vdots\\
\vdots &  &  & \ddots & h_{2N,2N+1}\\
h_{2N+1,1} & \cdots & \cdots & h_{2N+1,2N} & 0\end{array}\right)\;.\label{eq:QuadraticMajorana}\end{equation}
 The elements of the matrices ${{\mathcal{H}}^{\mathrm{Gauss}}}$
and $h$ are closely related because the theory is Gaussian -- there
will be modifications in the case of interacting systems. Note that
${\mathcal{H}}^{\mathrm{Gauss}}$ is an odd-dimensional Hermitian
antisymmetric matrix so it has an eigenvector with zero eigenvalue
and real components $\left(\alpha_{1},\alpha_{2},\dots,\alpha_{2N+1}\right)$
which corresponds to the zero mode $\mathcal{O}=\sum_{i}\alpha_{i}\gamma_{i}$.
Notice that it follows from the relations in Eq.~(\ref{eq:BasicDefintion})
that $\mathcal{O}^{\dagger}=\mathcal{O}$ and $\mathcal{O}^{2}=\sum_{i}\alpha_{i}^{2}\times\openone$.

Let us now introduce notation so as to arrive at the same ${{\mathcal{H}}^{\mathrm{Gauss}}}$
in a way that will be similar to the calculations for interacting
systems below. Matching the coefficients multiplying each operator
$\gamma_{i}$ on both sides of the equation $\left[H^{\mathrm{Gauss}},\mathcal{O}_{\lambda}\right]=\lambda\,\mathcal{O}_{\lambda}$
can be achieved easily if we think of the $\gamma_{i}$ as basis vectors
and define an inner product for operators $A$ and $B$ as $(A,B)\equiv{\rm Coeff}_{\openone}(A^{\dagger}B)$,
where\begin{equation}
{\rm Coeff}_{\openone}\left(z\,\openone+\sum_{i}\alpha_{i}\,\gamma_{i}+\sum_{i,j}\beta_{i,j}\,\gamma_{i}\gamma_{j}+\dots\right)\equiv z,\label{eq:Innerproduct}\end{equation}
 \textit{i.e.}, the function ${\rm Coeff}_{\openone}(\mathcal{Q})$
returns the coefficient proportional to the identity in the multinomial
expansion of the operator $\mathcal{Q}$. One can check that the inner
product is Hermitian, $(A,B)=(B,A)^{*}$ and it follows from the algebra
of the $\gamma_{i}$'s that the inner product gives $(\gamma_{i},\gamma_{j})=\delta_{i,j}$.

Armed with this inner product we then compute the matrix \begin{eqnarray}
{\mathcal{H}}_{ij}^{\mathrm{Gauss}} & = & \left(\gamma_{i},\left[H^{\mathrm{Gauss}},\gamma_{j}\right]\right)\nonumber \\
 & = & -\left(\gamma_{j},\left[H^{\mathrm{Gauss}},\gamma_{i}\right]\right)=-{\mathcal{H}}_{ji}^{\mathrm{Gauss}}\;,\label{eq:Antisymmetry}\end{eqnarray}
 where the last line follows by direct computation and the fact that
$h_{i,j}=-h_{j,i}\in\mathbb{R}$. Once again ${\mathcal{H}}_{ji}^{\mathrm{Gauss}}$
is given by Eq. (\ref{eq:QuadraticMajorana}) above. We thus arrive
once more at the result that zero modes can be determined from null
vectors of a linear eigenvector equation for a Hermitian anti-symmetric
matrix ${\mathcal{H}}_{ij}^{\mathrm{Gauss}}$ (of odd dimension).

\section{Interacting Hamiltonians}

\subsection{Quartic Hamiltonian }

We will consider a Hamiltonian given by: \begin{equation}
H^{\mathrm{Quart}}=i\sum_{i,j}h_{i,j}\;\gamma_{i}\gamma_{j}+\sum_{i,j,k,l}V_{i,j,k,l}\;\gamma_{i}\gamma_{j}\gamma_{k}\gamma_{l},\label{eq:QuarticHamiltonian}\end{equation}
 with $h_{i,j}$ a real and anti-symmetric matrix and $V_{i,j,k,l}$
real and antisymmetric under odd permutations of $i,j,k,l$ (we have
dropped an irrelevant constant that gives a state independent energy
shift). We will look for operators that commute with $H^{\mathrm{Quart}}$.
We will work with a vector space that is spanned by all linearly independent
Hermitian modes obtained from products of individual Majorana fermions
$\gamma_{i}$: \begin{eqnarray}
0\:\gamma: &  & \openone,\label{eq:BasisFermion}\\
1\:\gamma: &  & \gamma_{1},\gamma_{2},\gamma_{3},\dots,\gamma_{2N+1},\nonumber \\
2\:\gamma'\mathrm{s}: &  & i\gamma_{1}\gamma_{2},i\gamma_{1}\gamma_{3},\dots,i\gamma_{2N}\gamma_{2N+1},\nonumber \\
3\:\gamma'\mathrm{s}: &  & -i\gamma_{1}\gamma_{2}\gamma_{3},\dots,-i\gamma_{2N-1}\gamma_{2N}\gamma_{2N+1},\nonumber \\
\dots: &  & \dots\nonumber \\
2N+1\:\gamma'\mathrm{s}: &  & i^{\left(2N+1\right)N}\gamma_{1}\gamma_{2}\dots\gamma_{2N+1}\;.\nonumber \end{eqnarray}
 There are in total $\sum_{k=0}^{2N+1}{\binom{2N+1}{k}}=2^{2N+1}$
such operators, which we will denote by $\Upsilon_{a}$, for $a=1,\dots,2^{2N+1}$.
For each $a$ we define $n_{a}$ to be the number of $\gamma$'s in
the product $\Upsilon_{a}$, and we let $L(a)\equiv\{i_{1}(a),\dots,i_{n_{a}}(a)\}$
be the list of indices appearing in the product $\Upsilon_{a}$. With
this notation, one can write \begin{equation}
\Upsilon_{a}\equiv i\,^{{n_{a}(n_{a}-1)}/{2}}\;\gamma_{i_{1}(a)}\gamma_{i_{2}(a)}\dots\gamma_{i_{n_{a}}(a)}.\label{eq:Upsilon_a}\end{equation}
 The choice of phase factor guarantees that $\Upsilon_{a}=\Upsilon_{a}^{\dagger}$
and $\Upsilon_{a}^{2}=\openone$. Using Eq. (\ref{eq:Upsilon_a})
one verifies that, up to a phase, the product of two $\Upsilon_{a}$'s
gives a third: $\Upsilon_{a}\,\Upsilon_{b}=(i)^{s(a,b)}\,\Upsilon_{c}$,
where $c$ satisfies $L(c)=L(a)\cup L(b)\setminus L(a)\cap L(b)$
and \textbf{$s\left(a,b\right)\in\mathbb{N}$}. Without loss of generality,
we shall reserve the labels $a=1$ and $a=2^{2N+1}$ for the identity
and the total Majorana operators: $\Upsilon_{1}=\openone$ and $\Upsilon_{2^{2N+1}}=i^{\left(2N+1\right)N}\gamma_{1}\gamma_{2}\dots\gamma_{2N+1}\equiv\Upsilon_{\mathrm{Maj}}$.

We can now rewrite the Hamiltonian Eq.~(\ref{eq:QuarticHamiltonian})
as \begin{equation}
H^{\mathrm{Quart}}=\!\sum_{a|n(a)=2}h_{a}\;\Upsilon_{a}+\!\sum_{a|n(a)=4}V_{a}\;\Upsilon_{a}\;,\label{eq:QuarticHamiltonian2}\end{equation}
 for some coefficients $h_{a}$ , $V_{a}$ defined when $n(a)=2$
or 4, respectively, and $h_{a},\: V_{a}\in\mathbb{R}$. 
Below we will convert $H^{\mathrm{Quart}}$ into an operator acting
on the vector space spanned by the $\Upsilon_{a}$'s with the action
being given by the linear transformation where $H^{\mathrm{Quart}}$
acts by commutation: $\mathcal{O}\rightarrow\left[H^{\mathrm{Quart}},\,\mathcal{O}\right]$.
As a first step we extend the inner product given in Eq. (\ref{eq:Innerproduct})
above to the space spanned by $\Upsilon_{a}$ i.e. $(A,B)\equiv{\rm Coeff}_{\openone}(A^{\dagger}B)$.
One can check that the inner product is Hermitian, $(A,B)=(B,A)^{*}$
and the set $\Upsilon_{a}$ forms an orthonormal basis. Furthermore,
up to a multiplicative constant, we see that it is also given by the
usual trace inner product:\begin{equation}
\left(A,B\right)=\frac{1}{2^{2N+1}}{tr}\left(A^{\dagger}B\right).\label{eq:Innerproducttrace}\end{equation}
 Here, ${tr}$ is taken over the space spanned by $\Upsilon_{a}$.
Indeed this can be checked by noting that Eq. (\ref{eq:Innerproducttrace})
is linear, so it is sufficient to consider only terms of the form
$A=\Upsilon_{a},\: B=\Upsilon_{b}$. There are two possibilities:
1) $\Upsilon_{a}=\Upsilon_{b}$ in which case ${tr}\left(\Upsilon_{a}{}^{\dagger}\Upsilon_{b}\right)=2^{2N+1}$
(the dimension of the vector space) 2) $\Upsilon_{a}\neq\Upsilon_{b}$,
for which case ${tr}\left(\Upsilon_{a}{}^{\dagger}\Upsilon_{b}\right)=0$,
and Eq.~(\ref{eq:Innerproducttrace}) holds. We now compute the matrix
elements ${\mathcal{H}}_{ab}^{\mathrm{Quart}}$. Since $\left[H^{\mathrm{Quart}},\Upsilon_{b}\right]$
is an anti-Hermitian operator (or $i$ times a Hermitian operator)
all the matrix elements of ${\mathcal{H}}_{ab}^{\mathrm{Quart}}$
are imaginary. Now because $\left\{ \Upsilon_{b}\right\} $ is an
orthonormal set we may compute matrix elements by taking inner products:\begin{eqnarray}
{\mathcal{H}}_{ab}^{\mathrm{Quart}} & = & \left(\Upsilon_{a},\left[H^{\mathrm{Quart}},\Upsilon_{b}\right]\right)\label{eq:Oper-Quar}\\
 & = & \frac{1}{2^{2N+1}}{tr}\left(\Upsilon_{a}H^{\mathrm{Quart}}\Upsilon_{b}-\Upsilon_{a}\Upsilon_{b}H^{\mathrm{Quart}}\right)\nonumber \\
 & = & -\left(\Upsilon_{b},\left[H^{\mathrm{Quart}},\Upsilon_{a}\right]\right)=-{\mathcal{H}}_{ba}^{\mathrm{Quart}}\;,\nonumber \end{eqnarray}
 so ${\mathcal{H}}_{ab}^{\mathrm{Quart}}$ is antisymmetric. The equality
in the last line of Eq.~(\ref{eq:Oper-Quar}) comes from the cyclic
property of trace. Therefore we arrive at a Hermitian anti-symmetric
matrix ${\mathcal{H}}^{\mathrm{Quart}}$. So far, this matrix has
dimension $2^{2N+1}\times2^{2N+1}$, which is even. However, one can
break this matrix into four block-diagonal pieces. First, because
$H^{\mathrm{Quart}}$ contains only even $\Upsilon_{c}$, that is
with $n_{c}$ even, sectors with opposite parity are not mixed by
${\mathcal{H}}_{ab}^{\mathrm{Quart}}$, so necessarily $n_{a}\equiv n_{b}$
mod 2. Therefore we break ${\mathcal{H}}^{\mathrm{Quart}}$ into blocks
acting on the fermionic and bosonic $\left\{ \Upsilon_{a}\right\} $,
each block a $2^{2N}\times2^{2N}$ matrix. Second, notice that both
the identity and the total Majorana operator commute trivially with
$H^{\mathrm{Quart}}$, so they each reside in a $1\times1$ block.
The identity is in the even sector ($n_{1}=0$) and the total Majorana
operator is in the odd sector 
($n_{\mathrm{Maj}}=2N+1$). Therefore we have broken down ${\mathcal{H}}^{\mathrm{Quart}}$
into four odd-dimensional Hermitian and anti-symmetric block matrices:
there are four operators that commute with the Hamiltonian $H^{\mathrm{Quart}}$,
or zero mode solutions. They are, in the even block, the trivial identity
$\Upsilon_{1}=\openone$ and the Hamiltonian $H^{\mathrm{Quart}}$
proper, and in the odd sector the total Majorana operator $\Upsilon_{\mathrm{Maj}}$
\cite{Akhmerov} and \textit{another non-trivial solution} $\mathcal{O}=\sum_{a}\alpha_{a}\,\Upsilon_{a}$,
with $\alpha_{a}$ solutions of $\sum_{b}{\mathcal{H}}_{ab}^{\mathrm{Quart}}\,\alpha_{b}=0$.

\subsection{Generic Fermionic Hamiltonians}

Let us allow for arbitrarily high order interactions. That is we will
consider Hamiltonians of the form $H^{\mathrm{Gen}}=i\sum h_{i,j}\,\gamma_{i}\gamma_{j}+\sum_{i,j,k,l}V_{i,j,k,l}\,\gamma_{i}\gamma_{j}\gamma_{k}\gamma_{l}+i\sum_{i,j,k,l,m,n}Q_{i,j,k,l,m,n}\,\gamma_{i}\gamma_{j}\gamma_{k}\gamma_{l}\gamma_{m}\gamma_{n}+\dots$,
which may also be expressed as \begin{equation}
H^{\mathrm{Gen}}=\sum_{a|n(a)=2}\!\! h_{a}\;\Upsilon_{a}+\!\!\sum_{a|n(a)=4}\!\! V_{a}\;\Upsilon_{a}+\!\!\sum_{a|n(a)=6}\!\! Q_{a}\;\Upsilon_{a}+\dots,\label{eq:H-gen}\end{equation}
 where $h_{a},\: V_{a},\: Q_{a},\dots\in\mathbb{R}$. We can construct
the matrix ${\mathcal{H}}^{\mathrm{Gen}}$ similarly to what we did
above, it is still a Hermitian antisymmetric matrix. Nothing changes
in the argument, and the essence is that the Hamiltonian contains
only $\Upsilon_{c}$ with even $n_{c}$, and therefore one can break
${\mathcal{H}}^{\mathrm{Gen}}$ into four block diagonal pieces exactly
the same way we did for quartic Hamiltonians and obtain zero modes.

\subsection{Bosonic Modes }

We now partially extend our ideas to the case of an odd number of
Majorana fermions coupled to some bosonic modes. Our main limitation
is that in order to insure convergence, to have finite dimensional
matrices only -- we will {}``truncate'' the Hilbert space of the
bosonic modes to a finite number of states. More precisely we will
assume that the relevant Hilbert space for the bosons is $M$ dimensional
and labeled by the states $\left\{ \left|1\right\rangle ,\left|2\right\rangle ...\left|M\right\rangle \right\} $
\cite{key-8}. As such we may represent all boson operators by $M\times M$
Hermitian matrices. One can then write a Hamiltonian that generalizes
Eq.~(\ref{eq:H-gen}): \begin{eqnarray}
H^{\mathrm{Gen-Bose}} & = & \Theta^{M\times M}+\sum_{a|n(a)=2}\! h_{a}^{M\times M}\otimes\Upsilon_{a}+\nonumber \\
 & + & \!\!\!\!\sum_{a|n(a)=4}\!\!\!\! V_{a}^{M\times M}\otimes\Upsilon_{a}+\!\!\!\!\!\sum_{a|n(a)=6}\!\!\!\! Q_{a}^{M\times M}\otimes\Upsilon_{a}+\dots\nonumber \\
 & = & \sum_{a|n(a)\;{\rm even}}\;\;\sum_{p=1}^{M^{2}}\;\; W_{a,p}\;\Upsilon_{a}\otimes h_{p}\;,\label{eq:Gen-Bose}\end{eqnarray}
 with $\Theta^{M\times M},\, h_{a}^{M\times M},\, V_{a}^{M\times M},\, Q_{a}^{M\times M}$
Hermitian matrices and we expanded the bosonic $M\times M$ Hermitian
matrices into an orthonormal basis $\left\{ h_{1},\, h_{2},...h_{M^{2}}\right\} $,
with $(h_{p},h_{q})_{{\rm Bose}}=\delta_{pq}$. The inner product
is $(A,B)_{{\rm Bose}}\equiv\frac{1}{M}\,{tr}\left(A^{\dagger}B\right)$.
It is not too hard to see that this is a positive definite symmetric
form on the space of bosonic operators \cite{key-9}. Without loss
of generality, we take $h_{1}=\openone_{M\times M}$.

We can combine the operators in the fermionic and bosonic spaces and
define $\Omega_{a,p}\equiv\Upsilon_{a}\otimes h_{p}$, with the usual
tensor space inner product \cite{key-9}. These states are orthonormal
because $\left(\Omega_{a,q},\Omega_{b,q}\right)_{{\rm total}}\equiv\left(\Upsilon_{a},\Upsilon_{b}\right)\times(h_{p},h_{q})_{{\rm Bose}}=\delta_{a,b}\,\delta_{p,q}$.
We can also check that this is expressible as a trace: $\left(A,B\right)_{{\rm total}}=\frac{1}{2^{2N+1}}\frac{1}{M}{tr}\,\left(A^{\dagger}B\right)$.
Here the trace is over the total space spanned by $\Omega_{a,p}$.

Armed with these combined operators, we can show that there is an
exact zero mode in exactly the same way we have done in the previous
case. We need the matrix:\begin{eqnarray}
{\mathcal{H}}_{a,p;b,q}^{\mathrm{Gen-Bose}} & = & \left(\Omega_{a,p},\left[H^{\mathrm{Gen-Bose}},\Omega_{b,q}\right]\right)\label{eq:GenralizedAntisymmetric}\\
 & = & -\left(\Omega_{b,q},\left[H^{\mathrm{Gen-Bose}},\Omega_{a,p}\right]\right)=-{\mathcal{H}}_{b,q;a,p}^{\mathrm{Gen-Bose}}\;,\nonumber \end{eqnarray}
 which is Hermitian and anti-symmetric. The last equality in Eq. (\ref{eq:GenralizedAntisymmetric})
can be checked similarly to Eq. (\ref{eq:Oper-Quar}). We then break
${\mathcal{H}}_{a,p;b,q}^{\mathrm{Gen-Bose}}$ into even and odd block
diagonal spaces, as before. In this way, we find two zero modes in
the even sector, $\Upsilon_{1}\otimes h_{1}=\openone\otimes\openone_{M\times M}$,
and $H^{\mathrm{Gen-Bose}}$ proper, and two zero modes in the odd
sector, $\Upsilon_{\mathrm{Maj}}\otimes\openone_{M\times M}$ and
\textit{another non-trivial solution} $\mathcal{O}=\sum_{a,p}\alpha_{a,p}\,\Upsilon_{a}\otimes h_{p}$,
with $\alpha_{a,p}$ solutions of $\sum_{b,q}{\mathcal{H}}_{a,p;b,q}^{\mathrm{Quart}}\,\alpha_{b,q}=0$.

\begin{figure}
\begin{centering}
\includegraphics{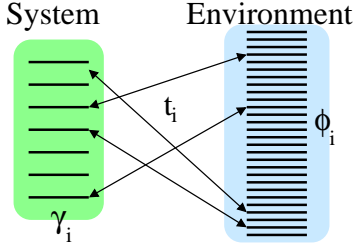} 
\par\end{centering}

\caption{\label{cap:tunneling} The system in tunneling contact with the environment.
The system is composed of CdGM states \cite{key-4}, while the environment
is everything else.}

\end{figure}

\vspace{0.1cm}

\section{Mode Counting and structure}

Let us count all zero modes in the system. We first start with the
Gaussian part of the theory, including bosons, and then later we add
the interactions. Consider a Hamiltonian given by: \begin{equation}
H^{\mathrm{Gauss}}=\sum_{m=1}^{M}E_{m}\left|m\right\rangle \left\langle m\right|+\frac{1}{2}\sum_{j=1}^{N}\epsilon_{j}\; i\gamma_{2j}\gamma_{2j+1}\;.\label{eq:GaussianDiagonalized}\end{equation}
 (Notice that $i\gamma_{2j}\gamma_{2j+1}=2\, c_{i}^{\dagger}c_{i}-1$.)
By inspection, there are $M\times2^{N}$ bosonic zero modes all given
by operators of the form $\mathcal{O}_{m,\{\theta_{j}\}}^{\mathrm{Bose}}\equiv\left|m\right\rangle \left\langle m\right|\otimes\prod_{j=1}^{N}\left(i\gamma_{2j}\gamma_{2j+1}\right)^{\theta_{j}}$
with $m=1,\dots,M$ and $\theta_{j}=0,1$ for $j=1,\dots,N$. There
are similarly $M\times2^{N}$ fermionic zero modes, simply given by
$\mathcal{O}_{n,\{\theta_{j}\}}^{\mathrm{Fermi}}\equiv\mathcal{O}_{n,\{\theta_{j}\}}^{\mathrm{Bose}}\;\gamma_{1}$.
These zero modes have a nice algebraic structure: 1) they are all
Hermitian, 2) appropriate linear combinations of them square to one:
$\left(\sum_{m}\mathcal{O}_{m,\{\theta_{j}\}}^{\mathrm{Fermi/Bose}}\right)^{2}=\openone$,
and 3) all zero modes commute: $\left[\mathcal{O}_{m,\{\theta_{j}\}}^{\mathrm{Fermi/Bose}},\mathcal{O}_{m',\{\theta'_{j}\}}^{\mathrm{Fermi/Bose}}\right]=0$.
As such any one of the fermionic modes (which squares to one), and
only one mode at a time, can be used for fermionic quantum computation
\cite{key-2}.

Let us now show that the number of zero modes and their commutation
relations do not change in the presence of weak interactions. To do
so, as a first step, consider the following family of Hamiltonians
$H^{\left\{ \delta\right\} }\equiv H^{\mathrm{{Gauss}}}+\sum_{m,\{\theta_{j}\}}\delta_{m,\{\theta_{j}\}}\,\mathcal{O}_{m,\{\theta_{j}\}}^{\mathrm{Bose}}$
with $\delta_{m,\{\theta_{j}\}}\in\mathbb{R}$, and we note that $\left\{ \delta_{m,\{\theta_{j}\}}\right\} \in\mathbb{R}^{M\times2^{N}}$.
It is not to hard to see that other then for points of accidental
degeneracy all zero modes of all Hamiltonians of the form $H^{\left\{ \delta\right\} }$
are given by $\mathcal{O}_{m,\{\theta_{j}\}}^{\mathrm{Fermi/Bose}}$.
As the next step, consider zero modes of Hamiltonians given by $H^{\left\{ \delta\right\} ,U}\equiv U^{\dagger}\, H^{\left\{ \delta\right\} }\, U$.
All the zero modes are now given by $U^{\dagger}\,\mathcal{O}_{m,\{\theta_{j}\}}^{\mathrm{Fermi/Bose}}\, U$,
and as such also satisfy conditions 1), 2), and 3) of the previous
paragraph. As before, exactly one appropriate mode from the fermionic
set can be used for quantum computation \cite{key-2}. To complete
the discussion of the counting and structure of the zero modes for
interacting systems, it remains for us to show that any Hamiltonian
with weak interactions can be written as a $H^{\left\{ \delta\right\} ,U}$.

To show this, we consider the map $\mathcal{F}:U\left(M^{2}\times2^{2N}\right)\oplus\mathbb{R}^{M\times2^{N}}\rightarrow\mathbb{R}^{M^{2}\times2^{2N}}$
given by $\mathcal{F}\left(U,\left\{ \delta_{m,\{\theta_{j}\}}\right\} \right)=U^{\dagger}H^{\left\{ \delta\right\} }U$.
It is enough to show that the image of $U\left(M^{2}\times2^{2N}\right)\oplus\mathbb{R}^{M\times2^{N}}$
contains a small open neighborhood of $H^{\mathrm{Gauss}}$. Indeed,
as any sufficiently weakly interacting Hamiltonian can be found in
a small neighborhood of a non-interacting one this would show that
$U^{\dagger}H^{\left\{ \delta\right\} }U$ is a representation of
all sufficiently weakly interacting Hamiltonians. By the implicit
function theorem it is enough to show that $d\mathcal{F}$ is a surjective
mapping onto $\mathbb{R}^{M^{2}\times2^{2N}}$. Now writing $U=e^{-i\widetilde{H}}$
we get $d\mathcal{F}\left(\widetilde{H},\left\{ \delta_{m,\{\theta_{j}\}}\right\} \right)=i\left[\widetilde{H},\, H^{\mathrm{Gauss}}\right]+\sum_{m,\{\theta_{j}\}}\delta_{m,\{\theta_{j}\}}\,\mathcal{O}_{m,\{\theta_{j}\}}^{\mathrm{Bose}}$.
{}From this we see that all the zero modes are explicitly in the
image of $d\mathcal{F}$. Since the transformation $*\rightarrow i\left[*,\, H_{\left\{ n\right\} ,\left\{ \gamma_{j}\right\} }^{Gauss}\right]$
is an invertible linear operator when restricted to the space of all
non-zero modes, all non-zero modes are also in the image of $d\mathcal{F}$
as well. As such all of $\mathbb{R}^{M^{2}\times2^{2N}}$ is in the
image of $d\mathcal{F}$. This shows that up to conjugation by a unitary
transformation the structure of the zero modes is the same as in the
non-interacting case completing the proof.

\section{Comparison with previous work}

In Ref.~\cite{Akhmerov}, the fermion parity operator $\Upsilon_{\mathrm{Maj}}$
was discussed. This Majorana operator commutes with \textit{any} Hamiltonian,
since it is formed by the product of \textit{all} the operators $\gamma_{i}$.
This operator sits on its own $1\times1$ block of the matrix $\mathcal{H}$,
for all cases studied, including in our generalization that includes
bosons interacting with the fermionic modes.

In contrast, the other zero mode solutions found in the larger odd-dimensional
block of $\mathcal{H}$ \textit{do} depend on the form of the Hamiltonian.
There are $M\times2^{N}-1$ of them. Furthermore one of the modes
has a particularly simple structure $\mathcal{O}=e^{i\widetilde{H}}\sum_{i}\alpha_{i}\gamma_{i}\, e^{-i\widetilde{H}}$
which is continuously connected to the non interacting mode (consider
$\mathcal{O}_{t}=e^{it\widetilde{H}}\sum_{i}\alpha_{i}\gamma_{i}\, e^{-it\widetilde{H}}$).
This mode is different from the fermion parity mode \cite{Akhmerov}
and, as we shall see below, for weak interactions (small $\widetilde{H}$)
it is better protected from various forms of decoherence when the
system is coupled to a generic bath.

\section{Decoherence}

Consider the setup shown in Fig. (\ref{cap:tunneling}). We consider
a simple perturbing tunneling Hamiltonian of the form: $\Delta H=i\sum_{i}t_{i}\gamma_{i}\eta_{i}$,
with $t_{i}\in\mathbb{R}$. Here $\eta_{i}$ refer to Hermitian fermionic
modes relevant to the environment. In previous works it was demonstrated
that $\left\langle \mathcal{O}\left(0\right)\mathcal{O}\left(T\right)\right\rangle $
is a good measure of the coherence of a qubit composed of localized
Majorana modes \cite{key-1}. Here $\mathcal{O}$ is an operator used
to encode the qubit, and we will assume that the qubit and environment
start uncorrelated. By Taylor expanding $e^{iT\Delta H}$ and keeping
only leading order terms we obtain $\left\langle \mathcal{O}\left(0\right)\mathcal{O}\left(T\right)\right\rangle =$
\begin{equation}
\begin{array}{l}
1-\frac{1}{2}T^{2}\sum_{i,j}t_{i}t_{j}\left\{ \left\langle \eta_{i}\eta_{j}\right\rangle \times\left\{ \left\langle \mathcal{O}\gamma_{i}\gamma_{j}\mathcal{O}\right\rangle +\left\langle \mathcal{O}\gamma_{i}\mathcal{O}\gamma_{j}\right\rangle \right\} \right.\\
\left.\qquad\qquad\qquad\quad+\left\langle \eta_{j}\eta_{i}\right\rangle \times\left\{ \left\langle \mathcal{O}\gamma_{j}\mathcal{O}\gamma_{i}\right\rangle +\left\langle \mathcal{O}^{2}\gamma_{j}\gamma_{i}\right\rangle \right\} \right\} .\end{array}\label{eq:FinalCorrelation}\end{equation}
 We can understand how this expression scales for various operators,
in particular for $\mathcal{O}=\Upsilon_{a}$, $n_{a}$ odd, we get
that $\left\langle \Upsilon_{a}\left(0\right)\Upsilon_{a}\left(T\right)\right\rangle =1-2T^{2}\sum_{i\in L\left(a\right)}t_{i}^{2}\left\langle \eta_{i}^{2}\right\rangle _{\mathrm{Env}}$.
Since $t_{i}^{2}\left\langle \eta_{i}^{2}\right\rangle _{\mathrm{Env}}\geq0$,
operators with larger $n_{a}$ decohere more quickly, at least for
short times. This indicates enhanced stability for operators that
are similar to single Majorana fermions, like the new zero modes presented
here.

\section{Braiding}

\subsection{Quadratic Hamiltonian}

As a warm up we will start with the case of quadratic Hamiltonians.
We would focus on the holomony under the exchange of vortices labeled
by 1 and 2. We would like to consider the idealized case of several
sets of fermionic zero modes $\left\{ \mathcal{O}_{m\left\{ \theta_{J}\right\} }^{Fermi,l}\right\} $,
of the form $\mathcal{O}_{m,\{\theta_{j}\}}^{Fermi,l}\equiv\left|m_{l}\right\rangle \left\langle m_{l}\right|\otimes\prod_{j=1}^{N}\left(i\gamma_{2j}^{l}\gamma_{2j+1}^{l}\right)^{\theta_{j}}\cdot\gamma_{1}^{l}$,
each set corresponding to its own individual finite environment, vortex.
The sets are labeled by $\ell$. We further assume that the individual
environments do not interact with the rest of the system. Since holomony
is given by a unitary transformation it preserves product structure:
$U_{Hol}^{\dagger}A\cdot BU_{Hol}=U_{Hol}^{\dagger}AU_{Hol}^{\dagger}\cdot U_{Hol}BU_{Hol}$.
As such it is enough to consider the holomony of single particle modes
$\left|m_{l}\right\rangle \left\langle m_{l}\right|$ and $\gamma_{i}^{l}$.
We start with $\left|m_{l}\right\rangle \left\langle m_{l}\right|$.
Since holomony preserves energy ordering, assuming no degeneracies,
under braiding $\left|m_{1}\right\rangle \rightarrow e^{i\theta_{m}}\left|m_{2}\right\rangle $
and $\left|m_{2}\right\rangle \rightarrow e^{i\overline{\theta}_{m}}\left|m_{1}\right\rangle $,
so overall $\left|m_{1}\right\rangle \left\langle m_{1}\right|\rightarrow\left|m_{2}\right\rangle \left\langle m_{2}\right|$
and $\left|m_{2}\right\rangle \left\langle m_{2}\right|\rightarrow\left|m_{1}\right\rangle \left\langle m_{1}\right|$.
Similarly following Ivanov \cite{key-14} we can work out the holomony
for the Majorana modes. We know that under a change of superconducting
phase by $\varphi$ the Majorana modes transform as $\gamma_{i}^{l}=\left(\begin{array}{c}
u_{i}^{l}\\
v_{i}^{l}\end{array}\right)\rightarrow\left(\begin{array}{c}
e^{i\varphi/2}u_{i}^{l}\\
e^{-i\varphi/2}v_{i}^{l}\end{array}\right)$. Since there is a change by $2\pi$ of the superconducting phase
when winding around a vortex and given that vortex two winds around
vortex one under braiding, we see that $\gamma_{i}^{1}\rightarrow\gamma_{i}^{2}$
and $\gamma_{i}^{2}\rightarrow-\gamma_{i}^{1}$. Combining we get
that \cite{key-3}: 

\begin{eqnarray}
\mathcal{O}_{m,\left\{ \theta_{j}\right\} }^{Fermi,1} & \rightarrow & \mathcal{O}_{m,\left\{ \theta_{j}\right\} }^{Fermi,2}\label{eq:Excahnge}\\
\mathcal{O}_{m,\left\{ \theta_{j}\right\} }^{Fermi,2} & \rightarrow & -\mathcal{O}_{m,\left\{ \theta_{j}\right\} }^{Fermi,1}.\nonumber \end{eqnarray}

We have reproduced Ising braiding statics.

\subsection{Generic Hamiltonians}

We would like to extend the derivation of Eq. (\ref{eq:Excahnge})
to the case of interacting modes. To do so we note that the many body
holomony for interacting zero modes is the same as the one body holomony
plus the effect of an additional Hamiltonian \cite{key-7,key-5,key-6}.
This Hamiltonian has matrix elements only between states of degenerate
energy for the instanteneous Hamiltonian of the system. For example
in the ground state manifold it is given by $H_{\Omega,\Omega'}^{Hol}=i\left\langle \Omega\right|\frac{d}{dt}\left|\Omega'\right\rangle $.
Here $\left|\Omega\right\rangle $ and $\left|\Omega'\right\rangle $
are instantaneous zero energy eigenkets. Similarly for other instanteneous
degenerate eigenkets. This Hamiltonian, which we shall not explicitly
compute, corresponds within the Heisenberg picture to an effective
evolution of the operators $U_{l}^{\dagger}\mathcal{O}_{m,\left\{ \theta_{j}\right\} }^{Fermi,l}U_{l}$.
This evolution is given by a unitary transformation generated by the
effective Hamiltonian $U_{l}^{\dagger}\mathcal{O}_{m,\left\{ \theta_{j}\right\} }^{Fermi,l}U_{l}\rightarrow P_{U}\left[H^{Hol},U_{l}^{\dagger}\mathcal{O}_{m,\left\{ \theta_{j}\right\} }^{Fermi,l}U_{l}\right]$,
where $P_{U}$ is the projector onto the space of zero modes (operators
in the manifold spanned by $U_{l}^{\dagger}\mathcal{O}_{m,\left\{ \theta_{j}\right\} }^{Fermi,l}U_{l}$).
Now we claim that for any Hamiltonian, in particular the holomony
Hamiltonian, $P_{U}\left[H,U_{l}^{\dagger}\mathcal{O}_{m,\left\{ \theta_{j}\right\} }^{Fermi,l}U_{l}\right]=0$.
We first note that: $P_{U}=\sum_{m,\left\{ \theta_{j}\right\} }\left|U_{l}^{\dagger}\mathcal{O}_{m,\left\{ \theta_{j}\right\} }^{Fermi,l}U_{l}\right\rangle \left\langle U_{l}^{\dagger}\mathcal{O}_{m,\left\{ \theta_{j}\right\} }^{Fermi,l}U_{l}\right|$,
so its enough to show that $\left(U_{l}^{\dagger}\mathcal{O}_{m,\left\{ \theta_{j}\right\} }^{Fermi,l}U_{l},\left[H,U_{l}^{\dagger}\mathcal{O}_{m,\left\{ \theta_{j}\right\} }^{Fermi,l}U_{l}\right]\right)=0$.
Now:\begin{equation}
\begin{array}{l}
\left(U_{l}^{\dagger}\mathcal{O}_{m,\left\{ \theta_{j}\right\} }^{Fermi,l}U_{l},\left[H,U_{l}^{\dagger}\mathcal{O}_{m,\left\{ \theta_{j}\right\} }^{Fermi,l}U_{l}\right]\right)\\
=tr\left\{ U_{l}^{\dagger}\mathcal{O}_{m,\left\{ \theta_{j}\right\} }^{Fermi,l}U_{l}\left[H,U_{l}^{\dagger}\mathcal{O}_{m,\left\{ \theta_{j}\right\} }^{Fermi,l}U_{l}\right]\right\} \\
=tr\left\{ \mathcal{O}_{m,\left\{ \theta_{j}\right\} }^{Fermi,l}\left[U_{l}HU_{l}^{\dagger},\mathcal{O}_{m,\left\{ \theta_{j}\right\} }^{Fermi,l}\right]\right\} .\end{array}\label{eq:Traces}\end{equation}
So its enough to prove $tr\left\{ \mathcal{O}_{m,\left\{ \theta_{j}\right\} }^{Fermi,l}\left[U_{l}HU_{l}^{\dagger},\mathcal{O}_{m,\left\{ \theta_{j}\right\} }^{Fermi,l}\right]\right\} =0$
for any Hamiltonian $U_{l}HU_{l}^{\dagger}$, e.g consider only the
non-interacting case. However by inspection $tr\left\{ \mathcal{O}_{m,\left\{ \theta_{j}\right\} }^{Fermi,l}\left[\left|m'\right\rangle \left\langle m'\right|\Upsilon_{a},\mathcal{O}_{m,\left\{ \theta_{j}\right\} }^{Fermi,l}\right]\right\} =0$.
So by taking linear combinations of terms of the form $\left|m'\right\rangle \left\langle m'\right|\Upsilon_{a}$
we see that any Hamiltonian is zero when acting on the space of zero
modes. As such the holomony reduces to the one given in Eq. (\ref{eq:Excahnge}).

\section{Conclusions}

We presented a systematic treatment of closed interacting systems
with an odd number of real fermions. This formulation allowed us to
find the zero mode solutions of interacting Hamiltonians, \textit{i.e.},
operators that commute with the many-body Hamiltonian. In addition
to the fermion parity operator that can be viewed as a constant of
the motion for \textit{any} Hamiltonian, we have found the solution
that connects continuously to the Majorana mode for non-interacting
systems as the interactions are switched off. These modes couple more
weakly than the fermion parity mode to an environment once the system
is opened up to an outside infinite bath~\cite{key-1}. Therefore,
the solutions that are continuously connected to the non-interacting
Majorana modes should lead to slower decay rates in the presence of
a bath. We have also verified that, under idealized conditions when
multiple such modes exist, they obey Ising like statistics under braiding.

This work was supported by NSF grant CCF-1116590.


\begin{thebibliography}{16}
\bibitem{Akhmerov} A. R. Akhmerov, Phys. Rev. B \textbf{82}, 020509
(2010).

\bibitem{key-2} S. B. Bravii and A. Y. Kitaev, Ann. Phys. \textbf{298},
210 (2002).

\bibitem{Jackiw-Rebbi} R. Jackiw and C. Rebbi, Phys.\ Rev.\ D \textbf{13},
3398 (1976).

\bibitem{Hou-Chamon-Mudry} C.-Y. Hou, C. Chamon, and C. Mudry, Phys.\ Rev.\ Lett.\ \textbf{98},
186809 (2007).

\bibitem{Jackiw-Rossi} R. Jackiw and P. Rossi, Nucl.\ Phys.\ B
\textbf{190}, 681 (1981).

\bibitem{Weinberg} E.~J.~Weinberg, Phys.\ Rev.\ D \textbf{24},
2669 (1981).

\bibitem{Fu-Kane} L.~Fu and C.~L.~Kane, \textit{Phys.\ Rev.\ Lett.}
\textbf{100}, 096407 (2008).

\bibitem{key-1} G. Goldstein, and C. Chamon, arXiv 1107.0288.

\bibitem{key-8} Spins are included in this formalism as a spin $S$
system corresponds to a $2S+1$ dimensional bosonic subspace.

\bibitem{key-9} S. Lang, \textit{Linear Algebra}, Springer Science
+ Business Media Inc., (1987).

\bibitem{key-4} C. Carroli, P. G. de Gennes, and J. Matricon, Phys.
Lett. \textbf{9}, 307 (1964).

\bibitem{key-14}D. A. Ivanov, Phys. Rev. Lett. \textbf{86}, 268 (2001).

\bibitem{key-3} For Bose modes the transformation is even simpler
$\mathcal{O}_{m,\left\{ \theta_{j}\right\} }^{Bose,1}\rightarrow\mathcal{O}_{m,\left\{ \theta_{j}\right\} }^{Bose,2},\:\mathcal{O}_{m,\left\{ \theta_{j}\right\} }^{Bose,2}\rightarrow\mathcal{O}_{m,\left\{ \theta_{j}\right\} }^{Bose,1}$.

\bibitem{key-7} J. E. Avron, R. Seiler and L. G. Yaffe, Commun. Math.
Phys. \textbf{110}, 33 (1987)

\bibitem{key-5} F. Wilczek and A. Zee, Phys. Rev. Lett. \textbf{52},
2111 (1984).

\bibitem{key-6} R. R. Aldinger, A. Bohm, M. Loewe, Found. of Phys.
Lett. \textbf{4}, 219 (1991).












\end{thebibliography}
\end{document}